\documentstyle[aps,preprint]{revtex}
\begin{document}
\title{Superconductivity and incommensurate spin fluctuations in a generalized $t-J$
model for the cuprates}
\author{C. D. Batista ($^*$), L. O. Manuel ($^{**}$), H. A. Ceccatto ($^{**}$) and
A. A. Aligia ($^*$)}
\address{($^*$) Centro At\'{o}mico Bariloche and Instituto Balseiro,\\
Comisi\'on Nacional de Energ\'{\i}a At\'omica,\\
8400 Bariloche, Argentina. \\
($^{**}$) Instituto de F\'{\i}sica de Rosario, \\
Boulevard 27 de Febrero 210 bis, 2000 Rosario, Argentina.}
\maketitle

\begin{abstract}
We consider the slave-fermion Schwinger-boson decomposition of an effective
model obtained through a systematic low-energy reduction of the three-band
Hubbard Hamiltonian. The model includes a three-site term $t^{\prime \prime }
$ similar to that obtained in the large-$U$ limit of the Hubbard model but
of opposite sign for realistic or large O-O hopping. For parameters close to
the most realistic ones for the cuprates, the mean-field solution exhibits $%
d+s$ superconductivity (predominantly $d_{x^2-y^2}$) with a dependence on
doping $x$ very similar to the experimentally observed. We also obtained
incommensurate peaks at wave vectors $\sim \pi (1,1\pm 2x)$ in the spin
structure factor, which also agree with experiment.
\end{abstract}


\noindent 75.10.J: Heisenberg and other quantized spin models.

\newpage

After the original derivation of the $t-J$ model by Zhang and Rice \cite{zr}%
, there has been much discussion about its validity \cite
{eme,che,rap,ali,ero}. At present it seems clear that while the low-energy
eigenstates contain $\sim 98\%$ of Zhang-Rice local singlets \cite{rap}
supporting the reduction of the Hilbert space, other terms in addition to
the nearest-neighbor hopping $t_1$ and exchange $J$ should be included to
accurately describe the low-energy physics of the original three-band
Hubbard model \cite{rap,ali,jef}. In particular, numerical comparison of the
energy levels of a Cu$_4$O$_8$ cluster has shown that it is necessary to
include the three-site correlated-hopping term $t^{\prime \prime }$ (see Eq.
(1)) to obtain a reasonable fit of the low-energy spectrum \cite{rap}.

While in the one-band Hubbard model $t^{\prime\prime} < 0$ (with the
notation of Eq. (1)), in the three-band Hubbard model $t^{\prime\prime}$
changes sign with increasing direct O-O hopping $t_{pp}$ and becomes
positive for realistic or large $t_{pp}$ \cite{ali,c}. Since most of the
terms in $t^{\prime\prime}$ have the effect of rotating $90^o$ a
nearest-neighbor singlet, one might expect that a negative (positive) $%
t^{\prime\prime}$ favors $s ~(d)$-wave pairing in the ground state. This is
in agreement with Monte Carlo calculations which show indications of $s$%
-wave \cite{dop} ($d$-wave \cite{kur}) superconductivity in the three-band
Hubbard model for $t_{pp}=0$ (sizeable $t_{pp}$). In addition, exact
diagonalization of a cluster of $4 \times 4$ unit cells has shown important
indications of a sudden large increase in the $d$-wave superconducting
correlations and off-diagonal long-range order for $t^{\prime\prime}>t^{%
\prime\prime}_c$, where $t^{\prime\prime}_c/t_1 \sim 0.1$ \cite{c}.
Simultaneously the wave function becomes similar to a simple short-range $%
(d+s)$-wave superconducting resonance-valence-bond (RVB) wave function. The
physics behind this is simple: the RVB wave function takes advantage of
both, the short-range magnetic correlations induced by $J$ and the hopping
of nearest-neighbor singlets $t^{\prime\prime}$.

However, the above mentioned studies suffer from finite-size effects and
either the temperature or the doping are not realistic. In particular, in
the $4 \times 4$ cluster, adding only two pairs of holes the system is
already in the overdoped region of real materials $x=0.25$, and a value of $%
t^{\prime\prime}$ nearly three times larger than the one obtained from the
mapping procedure was necessary to enter the RVB superconducting region \cite
{c}.

In this work we use a slave-fermion Schwinger-boson treatment \cite
{aue,shr,cha,nor} of the effective $t-t^{\prime\prime}-J$ model. While being
approximate, this method does not have the above mentioned shortcomings. 
In the undoped system the effective model reduces to the Heisenberg model,
whose ground-state properties are well accounted for by the Schwinger-boson
approximation \cite{aue,CGT}. For the doped system, we calculate the
doping-dependence ot both the superconducting gap and the wave vector of the
observed incommensurate peaks in the magnetic fluctuations \cite{cheo},
obtaining a good overall agreement with experiment. While previous numerical
work in the one-band Hubbard \cite{moreo} and generalized $t-J$ model \cite
{goo} obtained a deviation with doping $x$ of the maximum position in the
spin structure factor from $(\pi,\pi )$ towards $(\pi ,0)$, to our knowledge
no theory so far has been able to explain the fact that the deviation amount
is very near to $2\pi x$ \cite{cheo}.

The model can be written as \cite{c}: 
\begin{eqnarray}
H=-\sum_{ir\sigma }t_{ir}c_{r\sigma }^{\dagger }c_{i\sigma }+t^{\prime
\prime }\sum_{i,j\neq j^{\prime },\sigma }c_{j^{\prime }\sigma }^{\dagger
}c_{j\sigma }({\frac 12}-2{\bf S}_i\cdot {\bf S}_j)+{\frac J2}\sum_{ij\sigma
}({\bf S}_i\cdot {\bf S}_j-{\frac 14}n_in_j)~.
\end{eqnarray}
\noindent The first term contains hopping to first, second and third nearest
neighbors with parameters $t_1,~t_2,~t_3$ respectively. In the second,
correlated hopping term, $j$ and $j^{\prime }$ label the four nearest
neighbors of site $i$. The last term is the exchange term. The operator $%
c_{i\sigma }^{\dag }$ creates an electron on a vacuum where all sites carry
a Zhang-Rice singlet \cite{ali,c} and double occupancy is forbidden ($%
c_{i\sigma }^{\dag }n_i=0$). The parameters $t_1,~t^{\prime \prime }$ and $J$
are taken within the (narrow) band of possible values derived from a
realistic three-band Hubbard model \cite{ali,ero,c}. We take $t_1=1$ as the
unit of energy and $J=0.4,~t^{\prime \prime }=0.05$. Instead, $t_2$ and $t_3$
are included to account for effective second and third nearest-neighbor
hoppings which are dynamically generated in the bare $t-J$ model, and which
are not taken into account in our mean-field treatment. With $%
t_2=0.25,~t_3=0.18$, the quasiparticle dispersion of the bare $t-J$ model
calculated using the self-consistent Born approximation \cite{mar} or Monte
Carlo \cite{mc} is well reproduced by our approximation.

The constraint of no double occupancy is treated using the slave-fermion
Schwinger-boson decomposition, $c_{i\sigma}=b_{i\sigma} f_i^{\dagger},~
c^{\dagger}_{i\sigma} = f_i b^{\dagger}_{i\sigma}$ which, together with the
site-occupation constraint $\sum_{\sigma} b_{i\sigma}^{\dagger} b_{i \sigma}
+ f_i^{\dagger}f_i=1$, is a faithful representation of the original Fermi
algebra. In this case the spins can be written in terms of the Schwinger
bosons as ${\bf S}_i={\frac 1 2} b^{\dagger}_{i\sigma} {\bf \sigma}_{\alpha
\beta} b_{i \beta}$, while the slave fermion $f_i$ accounts for the charge
degrees of freedom. We prefer this decomposition since the Schwinger bosons
are known to produce a very good description of the long-range
antiferromagnetic order at half filling and the incommensurate magnetic
fluctuations upon doping \cite{cha,nor,CGT}. The proper account of these
magnetic fluctuations are essential to describe the low-energy physics of
the model, both at and near half filling. After the Hamiltonian is written
in terms of the new operators, the partition function is expressed as a
functional integral over coherent Bose and Fermi states. Furthermore, the
different terms of Eq. (1) are decoupled by means of Hubbard-Stratonovich
transformations according to the following scheme (indicated by a vertical
bar):

%

\begin{eqnarray}
H_t &\longrightarrow &-\sum_{ir\sigma }t_{ir}f_if_r^{\dagger }\ |\
b_{i\sigma }^{\dagger }b_{r\sigma },\qquad H_{t^{\prime \prime
}}\longrightarrow t^{\prime \prime }\sum_{i,j\neq j^{\prime }}\left(
f_i^{\dagger }f_j^{\dagger }\ |f_{j^{\prime }}f_i-\ f_j^{\dagger
}f_{j^{\prime }}\right) \ |\ {\hat{A}}_{ij^{\prime }}^{\dagger }\ |\ {\hat{A}%
}_{ij},  \nonumber \\
H_J &\longrightarrow &\frac J4\sum_{<ij>}\left( :{\hat{B}}_{ij}^{\dagger }\
|\ {\hat{B}}_{ij}:-{\hat{A}}_{ij}^{\dagger }\ |\ {\hat{A}}%
_{ij}-(1-x)^2\right) -\frac J2\sum_{<ij>}\left( f_i^{\dagger }f_j^{\dagger
}\ |\ f_jf_i\ |\ {\hat{A}}_{ij}^{\dagger }\ |\ {\hat{A}}_{ij}\right) 
\end{eqnarray}
The rationals behind these decompositions are as follows: For the hopping
terms we just decoupled bosons from fermions. In $H_J$ we took into account
the pairing effects over the holons $f$ of breaking magnetic bonds, and we
wrote the Schwinger-boson spin-spin interaction in terms of the two $SU(2)$
invariants ${\hat{B}}_{ij}^{\dagger }=\sum_\sigma b_{i\sigma }^{\dagger
}b_{j\sigma },~{\hat{A}}_{ij}=\sum_\sigma \sigma b_{i\sigma }b_{j{\bar{\sigma%
}}}$. These operators describe the magnetic fluctuations in the
ferromagnetic and antiferromagnetic channels respectively, and allow for a
proper treatment of the incommensurate order upon doping. Finally, $%
H_{t^{\prime \prime }}$ is conveniently written as boson and fermion pair
hopping, in terms of the antiferromagnetic singlet ${\hat{A}}_{ij}$ and
holon pairing operator. This decoupling highlights the connection between
superconductivity and short-range magnetic fluctuations observed in
numerical studies of the Hamiltonian \cite{c}.

According to the above decomposition of the interaction terms, in a
saddle-point evaluation of the partition function, the decoupling
Hubbard-Stratonovich fields become the following self-consistent order
parameters ($\eta$ connects nearest neighbors): 1) antiferromagnetic order: $%
A_{\eta}=\langle \sum_{\sigma} \sigma b_{i\sigma} b_{i+\eta,\sigma}\rangle $%
, 2) ferromagnetic order: $B_{\eta}=\langle \sum_{\sigma}
b^{\dagger}_{i\sigma} b_{i+\eta,\sigma} \rangle$, 3) superconducting order: $%
D_{\eta}=\langle f_i f_{i+\eta} \rangle$, and 4) nearest-neighbor hopping: $%
F_{\eta}=\langle f_i^{\dagger} f_{i+\eta}\rangle$. In ${\bf k}$-space the
saddle-point Hamiltonian can be written as:

%

\begin{eqnarray}
\label{hk}
H &=&\sum_{{\bf k},\sigma }\left[ \lambda _{{\bf k}}b_{{\bf k}\sigma
}^{\dagger }b_{{\bf k}\sigma }+R_{{\bf k}}\left( b_{{\bf k}\uparrow
}^{\dagger }b_{-{\bf k}{\downarrow }}^{\dagger }-{\rm H.c.}\right) \right]
+\sum_{{\bf k}}\left[ \epsilon _{{\bf k}}f_{{\bf k}}^{\dagger }f_{{\bf k}%
}+T_{{\bf k}}\left( f_{{\bf k}}^{\dagger }f_{-{\bf k}}^{\dagger }-f_{-{\bf k}%
}f_{{\bf k}}\right) \right]   \nonumber  \\
&&+N\left[ -\sum_\eta t_\eta F_\eta B_\eta -\frac J8\sum_\eta \left( \left|
B_\eta \right| ^2-\left| A_\eta \right| ^2+(1-x)^2\right) -3t^{\prime \prime
}\left| \sum_\eta D_\eta A_\eta \right| ^2\right.   \nonumber \\
&&\left. +3(t^{\prime \prime }+\frac J4)\sum_\eta \;\left| D_\eta A_\eta
\right| ^2+2t^{\prime \prime }\sum_{\eta \neq \eta ^{\prime }}F_{\eta -\eta
^{\prime }}A_\eta A_{\eta ^{\prime }}-\lambda -\mu x\right] .
\end{eqnarray}

\noindent
The subscript {\bf k} refers to the usual Fourier transform. However, we
have defined $b_{i\sigma }^{\dagger }=\frac 1{\sqrt{N}}\sum_{{\bf k}}b_{{\bf %
k}\sigma }^{\dagger }\;e^{i{\bf k}_\sigma r_i}$ with ${\bf k}_\sigma ={\bf k}%
+\sigma {\bf Q}/2$. The extra phase $\sigma {\bf Q.r}/2$ is required to
obtain the gapless (Goldstone) modes associated to the long-range magnetic
order at the proper ${\bf k=0,\pm Q}$ points. In Eq. (\ref{hk}) we have
defined:

%

\begin{eqnarray}
\lambda _{{\bf k}} &=&\lambda +\sum_\eta t_\eta F_\eta e^{i{\bf k}\eta }+%
\frac J4B_{{\bf k}},~R_{{\bf k}}=\frac J4A_{{\bf k}}+2t^{\prime \prime
}\sum_{\eta \neq \eta ^{\prime }}F_{\eta -\eta ^{\prime }}A_\eta e^{i{\bf k}%
\eta ^{\prime }}  \nonumber \\
&&-2t^{\prime \prime }D_{{\bf k}}\sum_\eta A_\eta D_\eta ^{*}+i(4t^{\prime
\prime }+J)(|D_x|^2A_x\sin k_x+|D_y|^2A_y\sin k_x),  \nonumber \\
\epsilon _{{\bf k}} &=&\lambda +\mu +\frac 1N\sum_{{\bf k}^{\prime }}t_{{\bf %
k-k^{\prime }}}B_{{\bf k}^{\prime }}-t^{\prime \prime }\left( |A_{{\bf k}%
}|^2-\sum_\eta A_\eta ^{*}A_\eta \right) ,  \nonumber \\
T_{{\bf k}} &=&t^{\prime \prime }A_k\sum_\eta A_{\eta}^{*}D_{\eta
}-2i(t^{\prime \prime }+\frac J4)(|A_x|^2D_x\sin
k_x+|A_y|^2D_y\sin k_y).
\end{eqnarray}

\noindent The Hamiltonian (\ref{hk}), quadratic in boson and fermion
operators, can be diagonalized in the standard way. The order parameters are
determined selfconsistently using a mesh of $200\times 200$ points for the
integrals in reciprocal space. Since our mean-field theory is not
variational because it unphysically enlarges the configuration space, we
cannot rely only on the energy-minimization criterion to seek for solutions
of the consistency equations. Consequently, based on previous numerical
studies of the $t-J$ model \cite{moreo,goo} we sought for solutions
corresponding to magnetic wavevectors $(\pi \pm Q,\pi )$ (which in some
parameter region can in fact have slightly higher energy than the
corresponding to a wavevectors along the diagonal). For these solutions, the
values of the bosonic order parameters as a function of doping $x$ are shown
at the left of Fig. 1. For $x\neq 0$, the Neel order turns continuously into
a spiral spin-density wave of increasing pitch, as a consequence of both the
finite value of the ferromagnetic order parameter $B_x$ and the decrease of
the antiferromagnetic parameters $A_x,A_y$ (mainly in the direction of the
spiral ${\hat{{\bf x}}}$). As expected, the dynamical susceptibility $\chi (%
{\bf q},\omega )$ is peaked at wave vectors $(\pi \pm Q,\pi )$. The
dependence of $Q$ on doping is very approximately linear, as represented at
the right of Fig. 1. The ratio $r=Q/(2\pi)\sim 1.2$, and is weakly dependent
on the parameters. It decreases for larger $J$. The experimentally observed
incommensurate magnetic fluctuations (with $r\sim 1$) \cite{cheo} are
consistent with microdomains of spiral density waves of both possible
directions (${\hat{{\bf x}}}$ and ${\hat{{\bf y}}}$).

The superconducting order parameter is given by 
\begin{equation}  \label{op}
\langle c^\dagger_{{\bf k} \uparrow} c^\dagger_{{\bf -k} \downarrow} -
c^\dagger_{{\bf k} \downarrow}c^\dagger_{{\bf -k} \uparrow}\rangle =2
(\Delta_x \cos k_x + \Delta_y \cos k_y)\ , \qquad \Delta_\eta=A_\eta D_\eta.
\end{equation}

\noindent The values of $\Delta_\eta $ are represented in Fig. 2. To show
the effect of $t^{\prime \prime }$ we also include in this figure the
corresponding result for $t^{\prime \prime }=0$ and $t^{\prime \prime
}=-0.05 $. In agreement with the experimental dependence of $T_c$ with
doping, the absolute value of the order parameter is maximum for a doping
slightly larger than $x=0.15$. As a consequence of the symmetry breaking
induced by the spiral spin arrangement, $|\Delta _x|\neq |\Delta _y|$ and
the order parameter has always mixed $s+d$ character. For $t^{\prime \prime
}\geq 0$, $\Delta _x$ and $\Delta _y$ have opposite signs, and the $d$-wave
component dominates. In agreement with previous numerical studies \cite{c},
a positive $t^{\prime \prime }$ increases considerably the magnitude of the
order parameter. 
The present results indicate that a small, realistic $t^{\prime \prime }$
leads to the proper order of magnitude of the superconducting gap ($\sim
(J+4t^{\prime \prime })\Delta _y\sim 0.005$eV for optimum doping). Instead,
a negative value of $t^{\prime \prime }$ (in the notation of Eq. (1))
suppresses superconductivity. In particular, for $t^{\prime \prime }<-0.04$
the signs of $\Delta _x$ and $\Delta _y$ are the same and, therefore, the $s$%
-wave component dominates the small superconducting signals. Since important
direct O-O hopping in the three-band Hubbard model leads to positive $%
t^{\prime \prime }$, while $t^{\prime \prime }=-J/4$ in the large-$U$ limit
of the Hubbard model, the present results agree with Monte Carlo results
which suggested $s$-wave \cite{dop} ($d$-wave \cite{kur}) superconductivity
in the three- band Hubbard model for $t_{pp}=0$ (sizeable $t_{pp}$) and the
absence of strong superconducting signals in the one-band Hubbard model.

In conclusion, using a slave-fermion Schwinger-boson mean-field
approximation applied to an effective model for the cuprates, we obtain a
superconducting state of mixed $(d+s)$-wave character with approximately
75\% of $d$-wave component. The order of magnitude of the gap and its
dependence with doping agree qualitatively with experiment. Moreover, he
symmetry of the order parameter agrees with different experiments which show
the existence of nodes, opposite signs along ${\hat{{\bf x}}}$ and ${\hat{%
{\bf y}}}$ directions, but also important tunneling currents along the $c$
axis in junctions with conventional $s$-wave superconductors \cite{dynes}.
In fact, Annet {\it et al.} have shown that a real combination of $s$ and $d$%
-wave components is consistent with most experiments \cite{ann}. However, it
is necessary to break the tetragonal symmetry to avoid a splitting of $T_c$.
We also obtain a spiral spin-density wave which provides this symmetry
breaking and is consistent with the incommensurate magnetic fluctuations
observed in superconducting La$_{2-x}$Sr$_x$CuO$_4$ \cite{cheo}. The spiral
spin density wave can be stabilized by the effective negative
next-nearest-neighbor hopping which appears in the holon dispersion relation 
\cite{cha,nor}, in agreement with numerical studies \cite{moreo,goo}.
Another possible explanation of the incommensurate fluctuations in terms of
dynamical stripe phases has been suggested \cite{tran,zaa}. A static version
of this phase consists of domain walls parallel to the ${\hat{{\bf x}}}$ (or 
${\hat{{\bf y}}}$) direction containing one hole every two Cu sites, and is
consistent with neutron-diffraction experiments in La$_{1.6-x}$Nd$_{0.4}$Sr$%
_x$CuO$_4$ with $x\sim 1/8$ \cite{tran}. However, domain-wall solutions of
Hubbard or $t-J$ models are always insulating \cite{schu,poil}, and as
stated by Schulz \cite{schu}, motion of domain walls is unlikely because
they are easily pinned by impurities and electron-lattice interactions.
Although further study is needed to establish the origin of the observed
incommensurate magnetic fluctuations, we find it encouraging that our
approach is consistent with them {\em and} with the essential features of
the superconducting state for the same set of realistic parameters.

Helpful discussions with J.J. Vicente are gratefully acknowledged. Three of
us (CDB,HAC,AAA) want to thank the hospitality of the International Centre
for Theoretical Physics, where part of this work was done. CDB and LOM are
supported by Consejo Nacional de Investigaciones Cient\'{\i}ficas y
T\'{e}cnicas (CONICET), Argentina. AAA and HAC are partially supported by
CONICET.


\section*{Figure Captions}

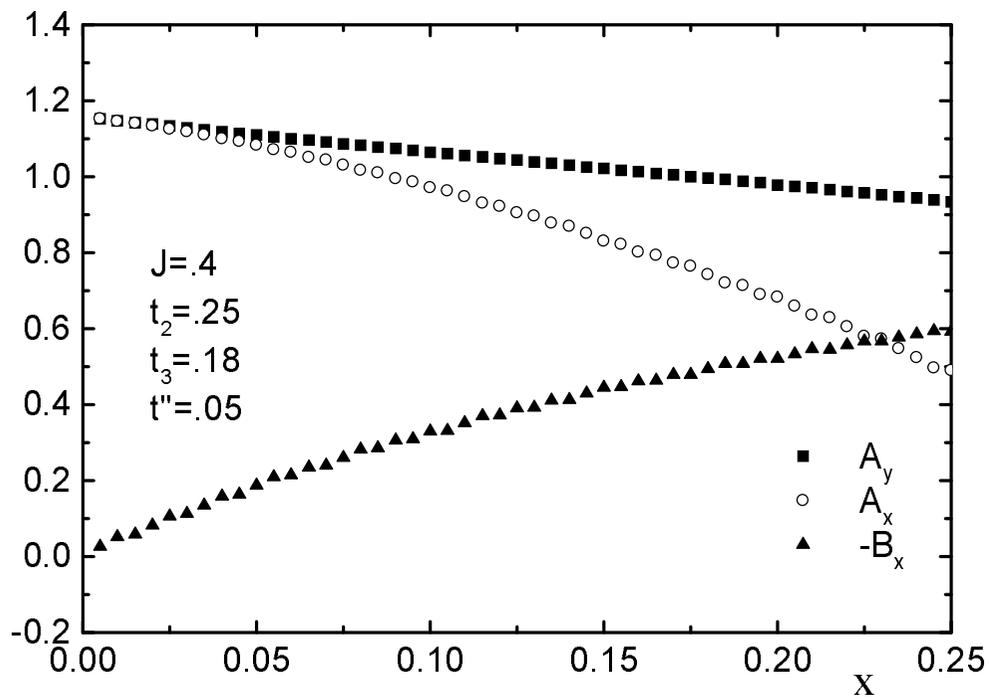
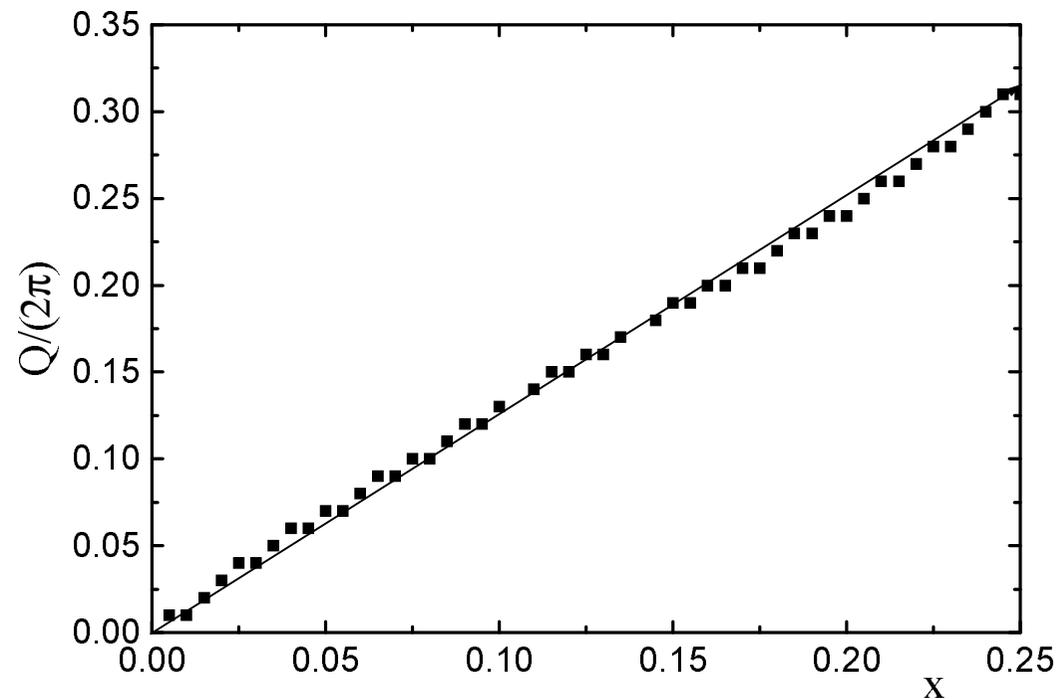
\noindent{\bf Fig.1:} Left: Order parameters $A_\eta, B_x$ as a function of
doping $x$ for one of the solutions (that with $B_y=0$) which minimize the
free energy (the other solution is equivalent by a rotation of $90^o$).
Right: position of the maxima of the dynamical susceptibility expressed as $%
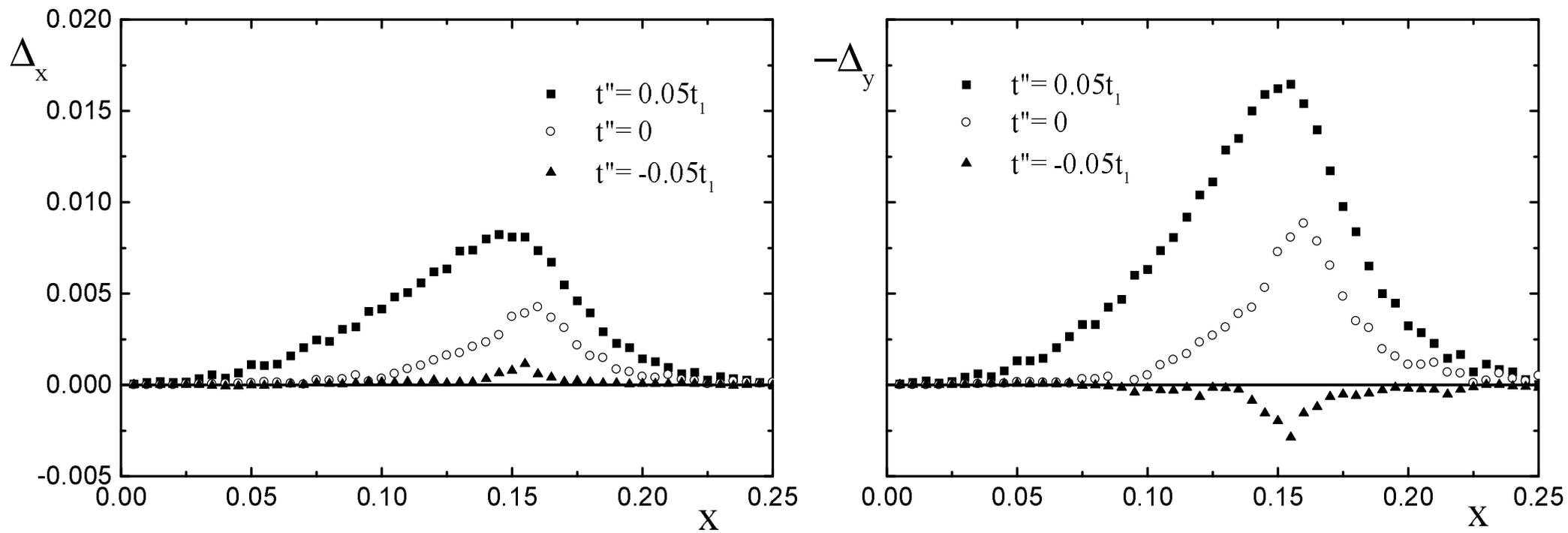
(\pi \pm Q,\pi)$ as a function of doping.\\\noindent{\bf Fig.2:}
Superconducting order parameter (see Eq. (\ref{op})) as a function of doping
for three values of $t^{\prime\prime}$ (from top to bottom, $%
t^{\prime\prime}=0.05,~0$ and -0.05).


\begin{references}
\bibitem{zr}  F.C. Zhang and T.M. Rice, {\it Phys. Rev. B} {\bf 37} (1988)
3759.

\bibitem{eme}  V.J. Emery and G. Reiter, {\it Phys. Rev.B} {\bf 38} (1988)
11938.

\bibitem{che}  C.X. Chen, H.B. Sch\"uttler and A.J. Fedro, {\it Phys. Rev.B} 
{\bf 41} (1990) 2581.

\bibitem{rap}  C.D. Batista and A.A. Aligia, {\it Phys. Rev. B} {\bf 48}
(1993) 4212; {\bf 49} (1994) 6436 (E).

\bibitem{ali}  A.A. Aligia, M.E. Sim\'{o}n, and C.D. Batista, {\it Phys.
Rev. B} {\bf 49} (1994) 13061.

\bibitem{ero}  J. Eroles, C.D. Batista and A.A. Aligia, {\it Physica C} {\bf %
261} (1996) 237.

\bibitem{jef}  L.F. Feiner, J.H. Jefferson, and R. Raimondi, {\it Phys. Rev.
B} {\bf 53} (1996) 8751; references therein.

\bibitem{c}  C.D. Batista and A.A. Aligia, {\it Physica C} {\bf 264} (1996)
319; {\it J. Low Temp. Phys.} {\bf 105} (1996) 591.

\bibitem{dop}  G. Dopf, A. Muramatsu, and W. Hanke, {\it Phys. Rev. Lett.} 
{\bf 68} (1992) 353.

\bibitem{kur}  K. Kuroki and H. Aoki, {\it Phys. Rev. Lett.} {\bf 76} (1996)
4401.

\bibitem{aue}  A. Auerbach and D. Arovas, {\it Phys. Rev. Lett.} {\bf 61}
(1988) 617.

\bibitem{shr}  B. Shraiman and E.D. Siggia, {\it Phys. Rev. Lett.} {\bf 62}
(1989) 1564; D.M. Frenkel and W. Hanke, {\it Phys. Rev. B} {\bf 42} (1990)
6711.

\bibitem{cha}  B. Chakravorty, N. Read, C.L. Kane and P.A. Lee, {\it Phys.
Rev. B} {\bf 42} (1990) 4819.

\bibitem{nor}  B. Normand and P.A. Lee, {\it Phys. Rev. B} {\bf 51} (1995)
15519.

\bibitem{CGT}  H.A. Ceccatto, C.J Gazza, and A.E. Trumper, {\it Phys. Rev. B}
{\bf 47} (1993) 12329.

\bibitem{cheo}  S.W. Cheong, G. Aeppli, T.E. Mason, H. Mook, S. Hayden, P.C.
Canfield, Z. Fisk, K.N. Clausen, and J.L. Martinez, {\it Phys. Rev. Lett.} 
{\bf 67} (1991) 1791.

\bibitem{moreo}  D. Duffy and A. Moreo, {\it Phys. Rev. B} {\bf 52} (1995)
15607.

\bibitem{goo}  R.J. Gooding, K.J.E. Vos, and P.W. Leung, {\it Phys. Rev. B} 
{\bf 49} (1994) 4119; {\it ibid} {\bf 50} (1994) 12866.

\bibitem{mar}  Z. Liu and E. Manousakis, {\it Phys. Rev. B} {\bf 44} (1991)
2414.

\bibitem{mc}  E. Dagotto, A. Nazarenko, and M. Boninsegni, {\it Phys. Rev.
Lett.} {\bf 73} (1994) 728.

\bibitem{dynes}  A.G. Sun, D.A. Gajewski, M.B. Maple, and R.C. Dynes, {\it %
Phys. Rev. Lett.} {\bf 72} (1994) 2267.

\bibitem{ann}  J. Annet, N. Goldenfeld, and A.J. Legget, to appear in {\it %
Physical Properties of High Temperature Superconductors}, Vol. 5, D.M.
Ginsberg (ed.) (World Scientific, Singapore, 1996).

\bibitem{tran}  J.M. Tranquada, B.J. Sternlieb, J.D. Axe, Y. Nakamura, and
S. Uchida, {\it Nature} {\bf 375} (1995) 561.

\bibitem{zaa}  J. Zaanen, M.L. Horbach, and W. van Saarloos, {\it Phys. Rev.
B} {\bf 53} (1996) 8671.

\bibitem{schu}  H.J. Schulz, {\it Phys. Rev. Lett.} {\bf 64} (1990) 1445.

\bibitem{poil}  D. Poilblanc and T.M. Rice, {\it Phys. Rev. B} {\bf 39}
(1989) 9749.
\end{references}
\end{document}